\documentclass[aps,pre,reprint,groupedaddress]{revtex4-2}

\usepackage{graphicx}
\usepackage{array,multirow}
\usepackage{amsmath}
\usepackage{color}
\usepackage{comment}

\begin{document}

\title{Traveling Fronts in Vibrated Polar Disks: \\ At the crossroad between Polar Ordering and Jamming}

\author{Caleb J. Anderson$^1$, Olivier Dauchot$^2$, Alberto Fernandez-Nieves$^{1,3,4}$}
\affiliation{$^1$Department of Condensed Matter Physics, University of Barcelona, 08028 Barcelona, Spain}
\affiliation{$^2$Laboratoire Gulliver, UMR CNRS 7083, ESPCI Paris, PSL University, 10, rue Vauquelin 75231 Paris de cedex 05, France}
\affiliation{$^3$ICREA-Institució Catalana de Recerca i Estudis Avançats, 08010 Barcelona, Spain}
\affiliation{$^4$Institute for Complex Systems (UBICS), University of Barcelona, 08028 Barcelona, Spain} 

\date{\today}

\begin{abstract}
We investigate experimentally the collective motion of polar vibrated disks in an annular geometry, varying both the packing fraction and the amplitude of the angular noise.
For low enough noise and large enough density an overall collective motion takes place along the tangential direction.
The spatial organization of the flow reveals the presence of polar bands of large density, as expected from the commonly accepted picture of the transition to collective motion in systems of aligning polar active particles.
However, in our case, the low density phase is also polar, consistent with what is observed when jamming takes place in a very high density flock.
Interestingly, while in that case the particles in the high density bands are arrested, resulting in an upstream propagation at a constant speed, in our case the bands travel downstream with a density-dependent speed.
We demonstrate from local measurements of the packing fraction, alignment, and flow speeds that the bands observed here result both from a polar ordering process and a motility induced phase separation mechanism.
\end{abstract}

\maketitle

\section{Introduction}

Collective-motion is a common feature in active matter comprised of self-propelled particles \cite{Vicsek2012,Marchetti2013}, whether these are synthetic, such as motile colloids \cite{Bricard2013,Morin2016,Geyer2018} and vibrated disks \cite{Deseigne2012, Briand2018}, or biological, such as humans \cite{Silverberg2013,Rio2018,Bain2019}, birds \cite{Ballerini2008,Nagy2010}, insects \cite{Bazazi2008,Latchininsky2013}, or cells \cite{Szabo2010}. 
Its prevalence across a wide variety of scales and its seeming independence from the precise nature of the underlying particles' pair-interactions, suggests that the important  physics in each of these systems is similar. 
In response to this idea, minimal models have been proposed and studied to quantify this emergent behavior.

Perhaps the most famous model, in which particles each move at constant speed and update their orientations at each time-step to noisily align with their nearby neighbors, was proposed by Vicsek et al. in 1995 \cite{Vicsek1995}. 
This model demonstrated that upon the reduction in the rotational noise in the alignment or upon an increase in the global density of the system, the particles spontaneously break symmetry from a state with no net momentum to one in which the particles are moving, on average, in the same direction.
Remarkably, the polar order is truly long range, even in two dimension, which would be forbidden by the Mermin-Wagner theorem in an equilibrium system~\cite{Mermin1966,Chaikin1995}.

Further work showed that the onset of this collective motion first manifests in the formation of polar traveling bands, which are regions in which the particles are dense and aligned, and therefore traveling together in the same direction.
These bands separate from, and coexist with, a dilute and completely disordered background, as evidenced in simulation \cite{Chate2008, Chate2020} and rationalized in a simplified version \cite{Caussin2014} of the large scale hydrodynamic equations of polar liquids, as described by Toner and Tu \cite{Toner1998, Toner2005, Toner2012}.
As the rotational noise in the system is decreased or the particle density is increased, the bands enlarge, become more numerous, and eventually merge together as the system approaches uniform collective motion.

Experimentally, the transition to collective motion was first reported in a system of aligning polar disks \cite{Deseigne2010}.
Careful measurements of the microscopic statistics of the particle displacements and collision rules \cite{Deseigne2012} allowed confirming the scenario associated to the transition in silico \cite{Weber2013}.
At the same time, a remarkable experiment conducted with tens of thousands of colloidal rollers characterized the details of the transition, from the dilute disordered gas, to the fully ordered polar state, further discussing all aspects of the resultant polar bands \cite{Bricard2013}.
The experiment was fully backed up by a kinetic theory analysis of the dynamics and the derivation of coarse-grained hydrodynamic equations.
This allowed a second set of experiments on sound propagation in the polar phase, confirming the predictions of a linear analysis of the Toner-Tu equations \cite{Geyer2018}.

Another striking non-equilibrium large-scale behavior that may be triggered by activity is clustering and phase separation in the absence of cohesive forces.
The tendency to form dense clusters arises from the combination of self-propulsion and excluded volume interactions.
The resultant clusters grow as self-propulsion or density is increased, eventually leading to a complete phase separation, the so-called Motility-Induced Phase Separation (MIPS) \cite{Cates2015} originally reported in the context of run-and tumble particles \cite{Tailleur2008}.
The interplay between the transition to collective motion and the motility induced phase separation leads to a complex phase diagram and remains a topic of intense research activity, using mostly numerical simulations~\cite{SeseSansa2018,MartinGomez2018,Paoluzzi2022}.
Experimentally, one aspect of this rich phase diagram was exemplified using again the system of colloidal rollers \cite{Geyer2019}.
At very high density, the crowding of the colloidal rollers lead to a motility induced phase separation inside the polar phase, in a way analogous to a traffic jam on a dense highway.
However, in contrast to cars, for the rolling colloids, their individual polarization is not intrinsic to the particles, but rather emerges as a result of a local symmetry breaking.
As a result, their jammed phase is disordered and thus fully arrested.
On the contrary, the vibrated disks have a well defined polarity, independently from their velocity.
Additionally, their alignment mechanism is robust, remaining operational at densities even close to close packing, as evidenced by the existence of a remarkable flowing-crystal phase~\cite{Briand2018}.

In this work, we take advantage of these properties of the vibrated polar disks to investigate experimentally the interplay of polar ordering and jamming in the case where density does not trivially suppress polar ordering.
To do so, we investigate collective motion in an annular geometry, varying both the global packing fraction and the amplitude of the angular noise.
For a wide range of parameter values, we observe polar bands of large density, as expected for the transition to collective motion, which however travel in a lower density \emph{polar phase}.
The bands travel downstream, with a wave speed faster than the mean flow velocity.
At moderate packing fraction these bands coexist with strong jams.
We demonstrate from local measurements of the packing fraction, alignment, and flow speed that both the bands and the jams observed here obey a local dynamical ``equation of state'', also called ``fundamental relation of traffic flows'' in other contexts~\cite{Nagatani2002}.
In the present case this relation results both from the polar ordering process, and the motility induced phase separation mechanism, corresponding to a bona-fide traffic jam, where the slow velocity dense phase remains polar and moves at a speed that decreases with density.

\section{Experimental Set-Up}
\begin{figure}[t!]
	\centering
	\includegraphics[width=3.2in]{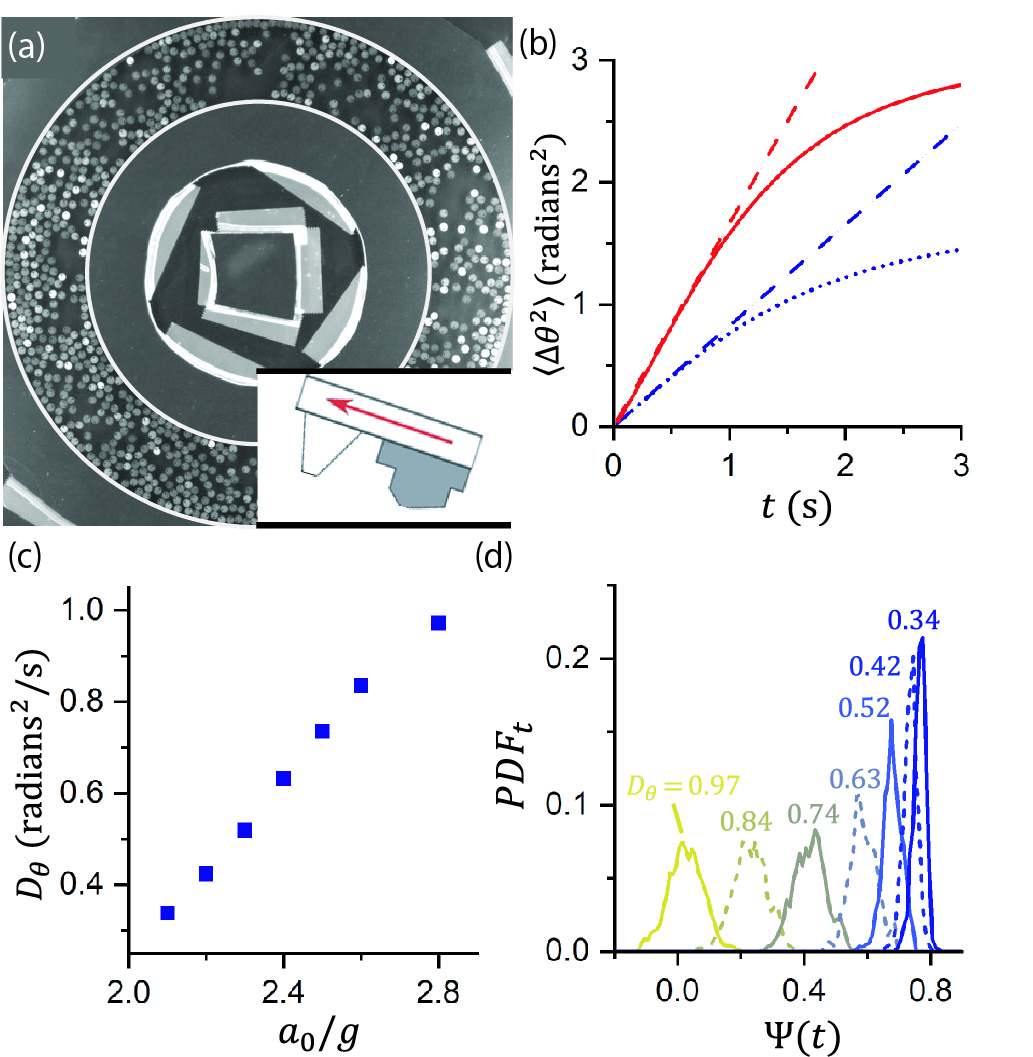}
	\caption{
	(a) A typical image of an experiment with \(\phi_0 =0.39\). The borders of the ring have been highlighted in white for clarity.
	(b) The mean squared change in the orientation of the disks as a function of time (solid and dotted lines) and corresponding short-time linear fits (dashed lines) for \(a_0/g=2.6\) (solid red line) and \(a_0/g=2.2\) (dotted blue line).
	(c) The rotational diffusion coefficient as a function of $a_0/g$.
	(d) Probability distribution functions for the instantaneous measurements of the order parameter for a sweep of various rotational diffusion values for \(\phi_0=0.39\). The numbers near the peak of each curve are \(D_\theta\) in inverse seconds.
	}
	\label{fig:SetUp}
\end{figure}

The system of vibrated polar disks has been described in detail elsewhere~\cite{Deseigne2012, Anderson2022}.
The disks are micro-machined, copper-beryllium cylinders with diameter \(d_0= (4.00 \pm 0.01)\)mm.
They each have one narrow metal off-center foot and a glued rubber skate located on opposite sides of the disk.
These {\sl feet} raise the disks to a height of \(2\)mm, and endow them with a polar axis, such that a vibration applied to the glass plate on which they stand causes them to undergo directed motion, as sketched in the inset in Fig. \ref{fig:SetUp}a.
The disks are confined from the top by a static lid, made of a thick glass plate.
The bottom and top plates are separated by \((2.40 \pm 0.05)\)mm.
The vibration is applied in the form of a well controlled vertical sinusoid displacement via an electromagnetic servo-controlled shaker (V455/6-PA1000L,LDS) coupled to a triaxial accelerometer (356B18, PCB Electronics).
The specificity of the set up for the present work is that the disks are confined in an annular geometry with inner diameter \(D_{in}=(180\pm2)\)mm and outer diameter \(D_{out}=(260\pm2)\)mm, as illustrated in~Fig. \ref{fig:SetUp}a, where we have highlighted the borders of the confining rings in white for clarity.

A digital camera acquires the motion of the disks at a frame rate of \(30\) images/s, from which we extract the position \({\bf r}_i\) and the orientation \(\theta_i\) of each disk in each frame. 
The velocity of each disk is then defined as \({\bf v}_i = ({\bf r}_i (t+\delta t) - {\bf r}_i (t))/\delta t\), with \(\delta t = 1 \)s and has components \((u_i,w_i)\) in polar coordinates. 
For choices of \(\delta t <<1 \)s, the motion of the disks is dominated by diffusive displacements and \({\bf v}_i \) does not represent well the orientation of the disks' directed motion. 
Additionally, \(\delta t = 1 \)s is approximately the limit of the ballistic regime of the mean squared displacement for the trial with highest orientational noise.

The disks' orientations experience an angular noise that we quantify by extracting the orientational diffusion constant, \(D_{\theta}\), from the mean squared angular deviation, \(\langle \Delta\theta^2 (t)\rangle\), as shown for two examples in Fig. \ref{fig:SetUp}b.
For large times, \(\langle \Delta\theta^2 (t)\rangle\) is not diffusive because of the annular confinement. 
Interactions between the disks and the walls of the annulus tend to align the disks' orientations along the wall, such that narrow annular confinement effectively forces the disks to flow either clockwise or counterclockwise.
Similarly, interactions between the disks tend to align individual disks with the direction of flow and decrease rotational diffusion.
We therefore extract \(D_{\theta}\) from the short times, where \(\langle (\Delta \theta)^2(t)\rangle=2D_\theta t\), using the slope of a linear fit of the first \(1/3\) s of the dynamics.
Here the dotted blue and solid red lines correspond to \(a_0/g=2.2\) and \(a_0/g=2.6\), respectively, with $a_0$ the maximum acceleration of the bottom plate and \(g\) the acceleration of gravity. 
Experimentally, we control the orientational diffusion constant by varying the amplitude of the vibrations at constant frequency, \(f=120\)Hz; this in turn affects $a_0$, further affecting $D_\theta$, as shown in Fig. \ref{fig:SetUp}c. 
For reference, a maximum acceleration of \(a_0=2g\) corresponds to a plate amplitude of 34$\mu$m. 
Manipulating the rotational noise of the particles this way also affects the average velocity of the disks in each trial.

\section{Approach to Collective Motion}
To examine the approach to collective motion, we perform measurements at a constant area fraction, \(\phi_0=0.39\), corresponding to 858 disks, and vary \(D_\theta\) over several trials. 
In all cases we find that the instantaneous average tangential speed of the disks \(u(t) = \frac{1}{N}\sum_i u_i \) is non zero.
Especially in the presence of large noise, however, the direction of the average flow switches back and forth between clockwise and counterclockwise.
These facts suggest our system is close or within the collective-motion phase, and additionally reflect its finite size.
At long times, the temporal average of \(u(t)\) should vanish, but for a given finite duration experiment, we always observe a dominant direction of the flow.
We conventionally set the sign of \(u\) to be positive when the disks are flowing in the direction of the time averaged flow.
We then define an order parameter related to the drift velocity of the disks around the ring: \(\Psi(t) = u(t)/\langle v \rangle\), where \(\langle v \rangle = \frac{1}{N}\sum_i \sqrt{u_i^2 + w_i^2} \) is the average speed of the disks throughout the experimental run.
Fig. \ref{fig:SetUp}d shows the probability distribution functions (PDFs) of instantaneous measurements of \(\Psi(t)\) for different \(a_0/g\). 
When \(D_\theta\) is relatively small, the disks align strongly and the measurements of \(\Psi(t)\) are narrowly distributed around values that indicate high polar order.
As \(D_\theta\) is increased, measurements of \(\Psi(t)\) skew towards lower order and also become more widely distributed.
The width of the distributions for high \(D_\theta\) indicates that the global order of the system fluctuates widely.
In particular, for the far left curve, corresponding to \(D_\theta =0.97 \text{s}^{-1}\), the PDF of \(\Psi(t)\) has a peak near 0, indicating very low order.
But note the peak is wide, including average tangential speeds in both clockwise and counterclockwise directions.
These unimodal distributions of the order parameter are typical of simulations with moderate numbers of active particles \cite{Nagy2007,Aldana2007,Chate2008}.

\section{Traveling Bands}
\begin{figure}[t]
	\centering
	\includegraphics[width=3.2in]{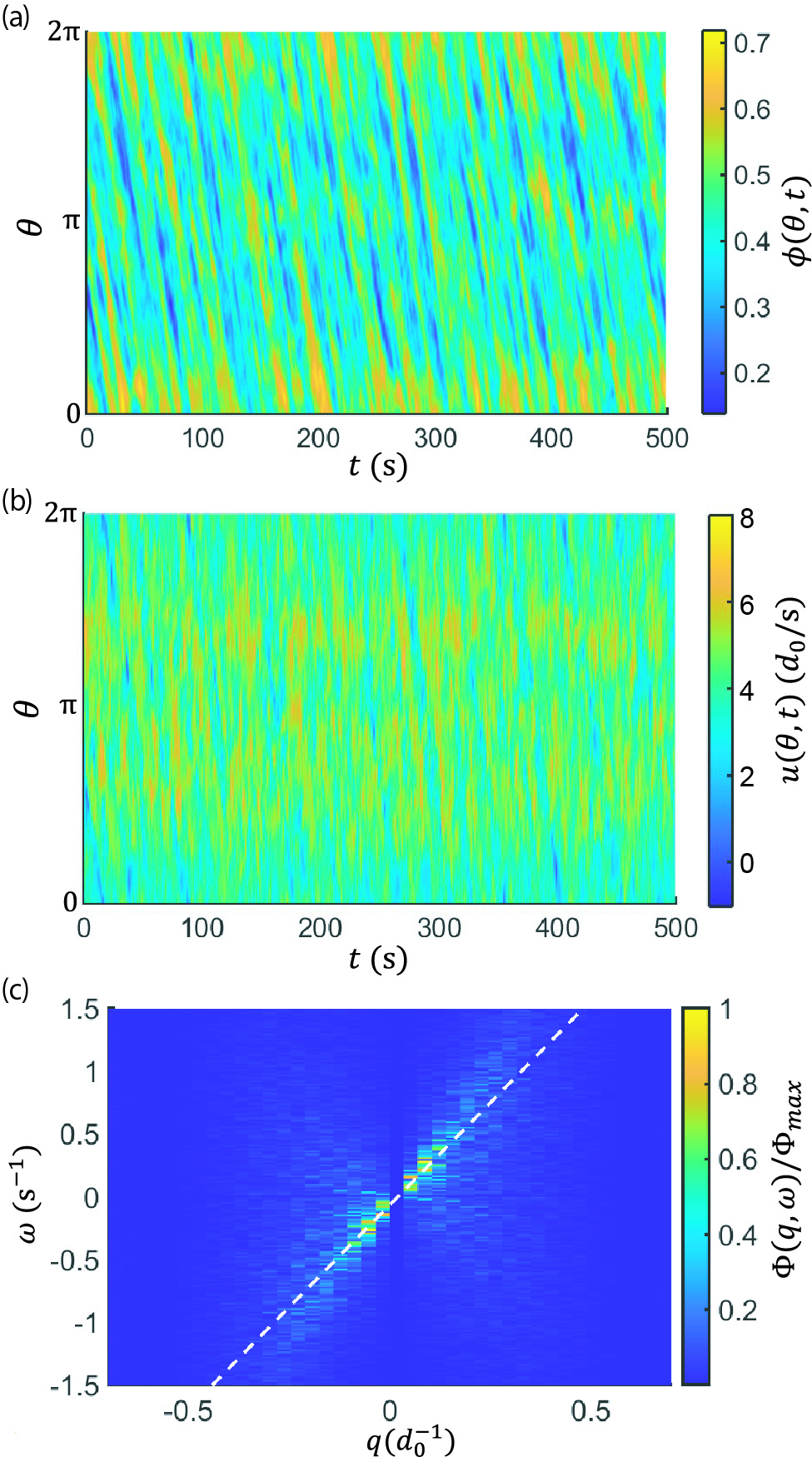}
	\caption{
	(a) The area fraction fraction (color scale) as a function of position around the ring and time for a trial with \(\phi_0 = 0.44\) and \(a_0/g=2.2\).  Fronts are visible as lighter patches propagating through a darker background.
	(b) The average tangential velocity of the disks as a function of space and time in the same trial.
	(c) The dispersion relation (color scale) for a trial with \(\phi_0=0.44\) and \(a_0/g=2.2\). The dotted white line is the linear fit around \((q,\omega)=(0,0)\); the slope of the line coresponds to \(c_0\).
	}
	\label{fig:Fronts}
\end{figure}

We also find that the disks undergoing collective motion around the ring spontaneously organize to form density fronts.
These can be clearly seen in the space-time plots of the local packing fraction \( \phi(\theta,t) \) (Fig. \ref{fig:Fronts}a), which we obtain by counting the number of particles in overlapping angular sectors of width \(21^o\), having an area of \(16.2\) cm$^2$. We also show in Fig. \ref{fig:Fronts}b the space-time plot of \( u(\theta,t) \), the angular velocity of the disks averaged in these same sectors, from which it is clear that \( u(\theta,t) \) never reaches zero or slightly-negative values, as it would if there were locally disordered regions.
We conclude that, unlike the traveling bands reported near the transition to collective motion in other systems or geometries, our ordered fronts propagate through an ordered background. Such fronts are observed whenever collective motion takes place - that is, for all trials except for the trial with \(D_\theta=0.97 \text{s}^{-1}\).

\begin{figure}[h]
	\centering
	\includegraphics[width=3.2in]{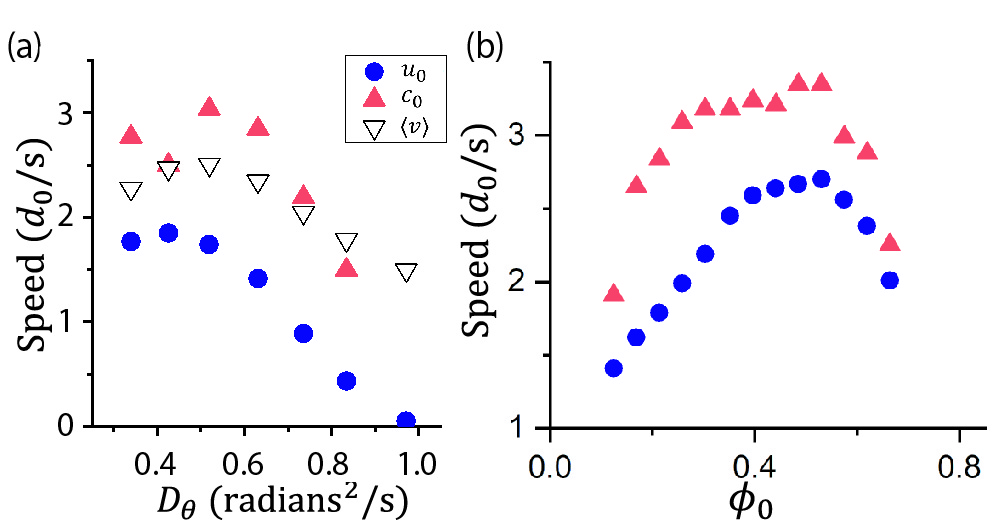}
	\caption{
	(a) The area average velocity of the disks in the direction tangentially around the ring (blue circles), the average speed of the fronts (red triangles), and the mean speed of the disks (down triangles) as a function of the disks' diffusion constant for \(\phi_0 = 0.39\). Note no value of front speed was found for the largest $D_\theta$.
	(b) The same two quantities as a funtion of the global area fraction.
	}
	\label{fig:Speed}
\end{figure}

To quantify the average speed of the traveling fronts, we compute the Fourrier transform, \(\Phi (q,\omega)= \Sigma_\theta \Sigma_t \phi (\theta,t) e^{-2\pi (q D_0 \theta+\omega t)}\), with \(D_0 = (D_{in}+D_{out})/2\), and focus on the maxima, which yield the dispersion relation \(\omega(q)\) of the propagating fronts. We then perform a linear fit of the dispersion relation around \((q,\omega) = (0,0)\), as shown in Fig. \ref{fig:Fronts}c, to obtain the speed of the dominant long-wavelength fluctuations in our system, \(c_0\), corresponding to the traveling fronts.
As shown in Fig. \ref{fig:Speed}a, the speed of the fronts decreases with increasing noise amplitude \(D_\theta\).
We also find that it is always larger than the average flow speed, averaged over the duration of the experiment \(u_0\), suggesting that the disks in the front are more aligned to flow tangentially to the ring, than the average disks in the trial.
Note that the vibration frequency of the plate has an effect on the average speed $<v>$ of the disks (see black triangles in Fig. \ref{fig:Speed}a), but the decrease in \(u_0\) and \(c_0\) with \(D_\theta\) is more pronounced.
Interestingly, when we vary the global area fraction \(\phi_0\), while holding \(a_0/g\) constant, therefore holding \(D_\theta=0.6 s^{-1}\), we find that the speed of the fronts also varies with \(\phi_0\), as shown in Fig. \ref{fig:Speed}b.
For moderate values of the packing fraction, the front speed increases with \(\phi_0\).
In contrast, for packing fraction \(\phi_0 > 0.55\), both the average flow speed and the front speed decrease.
This change in behavior hints at the fact that two competing mechanisms govern the flow speed in the disks.

\begin{figure*}
	\centering
	\includegraphics[width=5in]{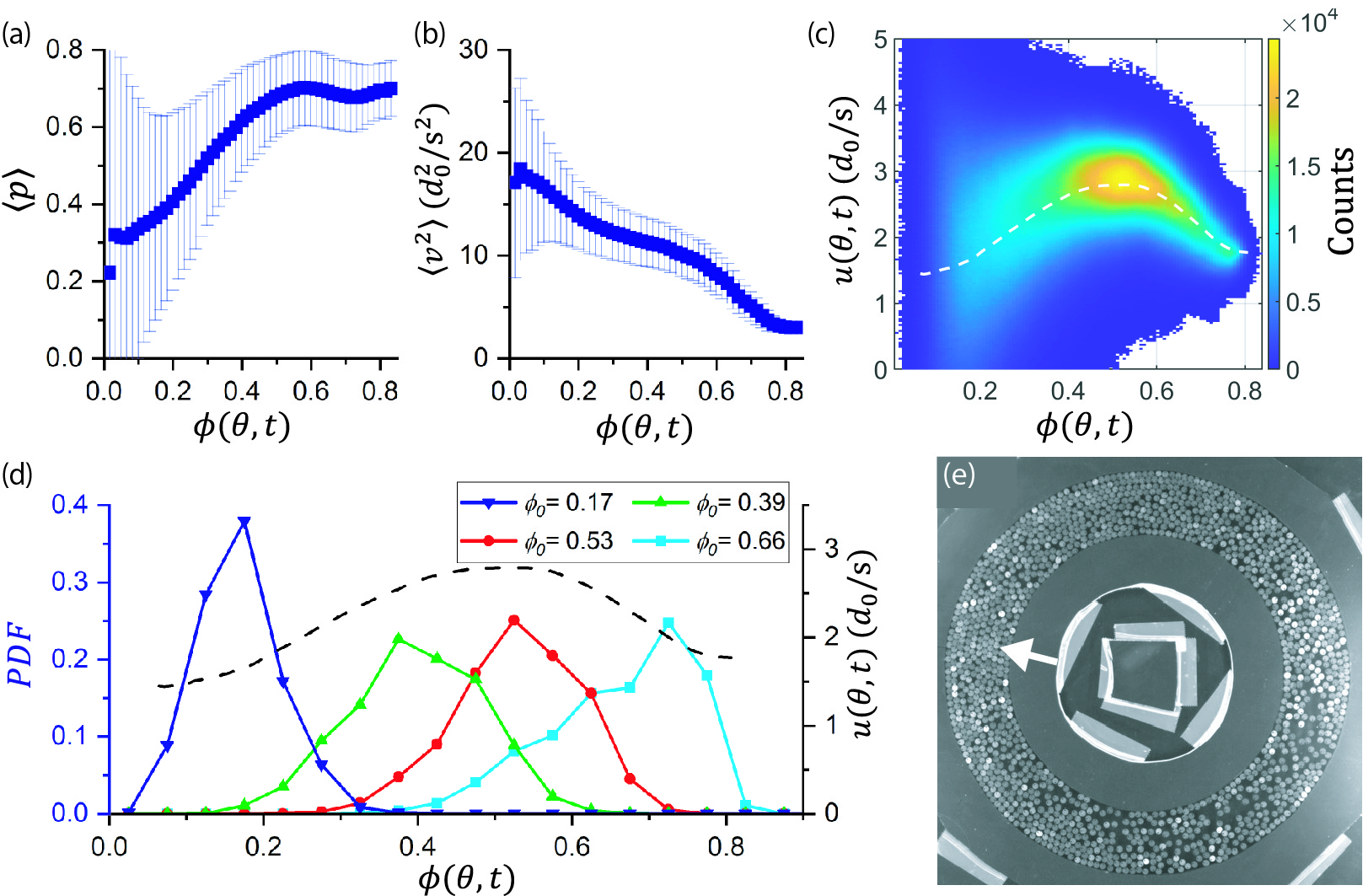}
	\caption{
	(a) The average alignment of the disks in a \(21^o\) bin as a function of the area fraction in that bin. The error bars represent the standard deviation of our measurements.
	(b) The average mean squared velocity of the disks in a bin as a function of the area fraction.
	(c) A 2D histogram of the total counts for each area fraction and local average velocity in the direction tangential to the ring over the course of the density sweep. The color scale indicates the total number of counts in each 2D bin.
	(d) The black dashed line is the average flow speed of the disks as a function of area fraction over the density sweep (left axis). The same curve is shown in panel (c) as a dashed white line. The solid lines with points are the PDFs of the area fraction measured in each of the \(21^o\) bins over the course of a single trial (left axis). From left to right, \(\phi_0=0.17, 0.39, 0.53, 0.66\).
	(e) An image of one of the long lived jams, indicated by the arrow, caused by the inverse dependence of the speed on density for high area fractions.
	}
	\label{fig:State}
\end{figure*}

\section{Local Equation of State}
We propose the non-monotonic behavior of both \(u_0\) and \(c_0\) is due to competing aligning and crowding effects. 
To test this idea, we measure ``local equations of state'' that relate the local averages of the alignment, mean squared velocity, and tangential speeds of the disks to the local area fraction.
We first measure the average alignment, \(p(\theta,t)\), defined as the average projection of the unit vector corresponding to \(\theta_i\) onto to the direction tangential to the annulus, and the mean squared velocity, \( v^2(\theta,t) \), of the disks in each of the instantaneous \(21^o\) averaging windows used to construct the above space-time diagrams. 
In the interest of obtaining better statistics and a broader range of local area fractions, \(\phi(\theta,t)\), we group together all measurements of \(p(\theta,t)\) and \( v^2(\theta,t) \) from the density sweep in Fig. \ref{fig:Speed}b and sort these measurements into bins by \(\phi(\theta,t)\). 
In doing so, we obtain parametric plots of the average alignment and the mean squared velocity as a function of the local packing fraction (Fig. \ref{fig:State}a,b).

We find that \(\langle p \rangle\) smoothly increases with \(\phi\), until it saturates around \(\phi(\theta,t)=0.55\).
At the same time, \(\langle v^2 \rangle\) monotonically decreases as a function of \(\phi(\theta,t)\), indicating that crowding effects are important at all of our experimentally studied densities.
These two effects combine to cause the trend in the local coarse grained flow speed of the disks, as can be seen in the 2D histogram of the measurements of \(\phi(\theta,t)\) and \(u(\theta,t)\) across the \(\phi_0\)-sweep at constant rotational noise in Fig. \ref{fig:State}c.
Here, the color scale indicates the number of observed simultaneous measurements of \(\phi(\theta,t)\) and \(u(\theta,t)\), and the dotted line shows the average \(u(\theta,t)\) as a function of \(\phi(\theta,t)\). 
At low packing fractions, the increase in alignment due to the increasing number of collisions more than compensates for the collisions' slowing effect, and \(\frac{du}{d\phi}>0\).
Conversely, for larger packing fraction \(\phi(\theta,t)>0.55\), when the alignment saturates, any increase in the local density tends to slow the disks down.

The traveling bands discussed in the previous section appear to be created by the first of these two effects.
At low packing fraction, a positive density fluctuation further polarizes the grains via the generic coupling between density and order that drives the transition to collective motion.
This high density, high order, region then travels more quickly, picking up the slower particles in front of it, becoming even denser.
This process continues until the local packing fraction is about \(0.55\), for which the typical flow speed is about \(u(\theta,t) \approx 3d_0/s\), or until a negative density fluctuation destroys the traveling band. 
Notice that this flow speed is closest to the speed of the dominant long-wavelength fluctuations in trials with global packing fractions \(0.28\leq\phi_0\leq.53\) in Fig. \ref{fig:Speed}b, further hinting that bands with \(\phi(\theta,t) \approx 0.55\) are the most stable.

At the largest packing fractions, jamming tends to become more important than alignment, such that positive density fluctuations now slow down the particles in that region.  
The mismatch in local speeds inside and outside the fluctuation again reinforces the the original fluctuation in a manner that is the essence of the MIPS mechanism.
The effect is also self-reinforcing, because the particles behind the slow-down are moving more quickly than those in the slow down, which further increases \(\phi(\theta,t)\) in the slow region.
This runaway effect continues, driving the local density very high. 
For example, see the perpetual jam formed in the trial with \(\phi_0=0.66\) indicated by the white arrow in Fig. \ref{fig:State}e. 
At these higher global packing fractions, the traveling bands detected by the dispersion relations in the previous sections are groups of particles that, as they break free from the leading end of the jam find themselves in the depleted region in front of the jam where they can propagate without further particles joining from behind the band.
 
In principle the growth of the large density fronts, whether it is driven by the aligning mechanism or the MIPS mechanism could take place for well separated packing fraction values, as it is the case for the colloidal roller experiments~\cite{Bricard2013,Geyer2019}.
However in the present case, the values of the effective parameters controlling the large scale properties of the system are such that both physics take place in the same range of values of these parameters.
Also, the system is not very large and local fluctuations can reach values that drives one or the other mechanism.
This is best illustrated when looking at the probability distribution of the local packing fractions for experimental runs conducted at a given nominal value \(\phi_0\) (see Fig.~\ref{fig:State}d). 
The experimental run with \(\phi_0=0.53\) is particularly illustrative.
Any fluctuation away from the mean density tends to slow the particles down and the macroscopic behavior of the system results from a subtle mix of the two mechanisms.

\section{Conclusions}
The selection mechanism of polar, or non polar, high density bands in systems of self propelled aligning particles results from complex non linear and stochastic dynamics~\cite{Caussin2014,Solon2015}.
Here we considered a system of self propelled disks, for which the alignment mechanism is robust to the slowing down induced by large density.
We observed a crossover between the traveling bands resulting from the first order nature of the transition to collective motion and jams produced by the slowing down of a high density system of self propelled agents. 
Also, the system size considered in our study facilitates large fluctuations, which favor the broadening of this crossover.
This experimental situation is illustrative of what is likely to take place in many dense active systems, including bacterial colonies, tissues, and traffic jams, calling for a deeper theoretical investigation. 
On the one hand both types of solutions were looked for and captured using large scale hydrodynamic models~\cite{Caussin2014, Geyer2019}.
It would be of interest to investigate the present crossover in this framework.
On the other hand, the important role of fluctuations call for a description in terms of stochastic hydrodynamics. 
We expect the diameter of the annulus to not strictly relate to the formation of the bands discussed here, as it does not enter the explanation for the bands' existence. 
Increasing the width of the annulus while maintaining the same density of disks would likely decrease the importance of fluctuations and could lead to phase separation.
However further work should be done to explore this further, as it is possible that the high curvature of the confining walls contributes to the magnitude of the density fluctuations in our system.
In a similar geometry, active brownian particles were shown to exhibit reentrant behavior, transitioning from collective motion to MIPS and back to collective motion, with increasing Peclet number \cite{Knippenberg2023}.
In this case, the presence of soft boundaries were instrumental to allow the particles to escape from MIPS at high Peclet number and transition to the collective-motion phase.
Since our confining walls are rigid, we do not expect to see this type of reentrant behavior in our disks, even if we were able to further reduce the disks' angular noise relative to their speed.
Overall, integrating the motility induced phase separation, within the polar ordering hydrodynamics seem to be a necessary step to capture the behavior of dense systems of aligning polar particles, even of simplest ones like the self propelled polar disks considered in the present work.

\begin{acknowledgments}
We thank financial support from MCIN/AEI/10.13039/501100011033/FEDER,UE (Grant No. PID2021-122369NB-100), the 2021 SGR 00450, and the FLAMEL (NSF DGE-1258425) and REU (NSF Grant GR10002751) programs. We are also thankful to Guillaume Briand for help in designing the experiments.
\end{acknowledgments}

\bibliography{references}

\end{document}